\documentclass[twocolumn,showpacs,preprintnumbers,amsmath,amssymb,pre]{revtex4}
\usepackage{graphicx}% Include figure files
\usepackage{dcolumn}% Align table columns on decimal point
\usepackage{bm}% bold math
\begin{document}
\title{Bulk and wetting phenomena in  a colloidal mixture of hard 
spheres and platelets}
\author{L. Harnau and S. Dietrich}
\affiliation{
         Max-Planck-Institut f\"ur Metallforschung,  
         Heisenbergstr.\ 3, D-70569 Stuttgart, Germany, 
         \\
         and Institut f\"ur Theoretische und Angewandte Physik, 
         Universit\"at Stuttgart, 
         Pfaffenwaldring 57, 
         D-70569 Stuttgart, Germany
	 }
\date{\today}
\begin{abstract}
Density functional theory is used to study binary colloidal fluids consisting of 
hard spheres and thin platelets in their bulk and near a planar hard wall. This 
system exhibits liquid-liquid coexistence of a phase that is rich in spheres 
(poor in platelets) and a phase that is poor in spheres (rich in platelets). 
For the mixture near a planar hard wall, we find that the phase rich in spheres 
wets the wall completely upon approaching the liquid demixing binodal from the 
sphere-poor phase, provided the concentration of the platelets is smaller
than a threshold value which marks a first-order wetting transition at 
coexistence. No layering transitions are found in contrast to recent studies 
on binary mixtures of spheres and non-adsorbing polymers or thin hard rods.
\end{abstract}

\pacs{64.70.Ja, 68.08.Bc, 82.70.Dd}
\maketitle

\section{Introduction}
Rich bulk phase diagrams involving colloidal gas, liquid, and solid phases
are found when non-adsorbing polymers or hard rodlike colloids are added as 
depletion agents to suspensions of colloidal spheres \cite{poon:96,adam:98}.
The chemical potential of the polymers or rods, with which one can tune their 
concentration, plays a role equivalent to that of the inverse temperature 
for a simple one-component substance characterized by a soft pair potential. 
Moreover, it has been shown theoretically \cite{brad:02b,roth:03} and by 
computer simulation \cite{dijk:02} that entropic depletion mechanisms lead to 
interesting wetting phenomena and to a rich surface phase behavior in 
colloidal mixtures of spheres and non-interacting polymers or rods. A wetting 
transition and layering transitions have been found when the mixtures are 
exposed to a hard wall. 

Recently the depletion potential between two hard spheres due to the presence 
of hard disclike colloids has been investigated \cite{piec:00,over:03,harn:04,over:04a}. 
Subsequently it has been shown within of a free-volume theory that depletion 
induced phase separation in a colloidal sphere-platelet mixture should occur at 
low platelet concentrations in systems now experimentally available \cite{over:04}. 
In view of the importance of such suspensions in biomedicine \cite{mans:02} and 
geophysics \cite{mait:00} we investigate in the present paper bulk and wetting 
phenomena of sphere-platelet mixtures using density functional theory. 
We demonstrate that the geometry-based density functional theory developed for binary 
mixture of hard spheres and thin rods \cite{schm:01a,schm:01b,schm:02,brad:02a}
can be consistently extended to the more challenging problem of hard spheres mixed 
with thin hard platelets (Sec.~II). Moreover, we study the bulk phase diagram 
(Sec.~III) and the wetting of the mixture at a planar hard wall by considering 
the platelets as thin and non-interacting regarding their mutual interactions 
(Sec.~IV). Our study provides a direct comparison of the bulk and wetting 
properties of binary sphere-platelet and sphere-rod mixtures.

\section{Density functional and fundamental measure theory}  
\begin{figure}[t]
\begin{center}
\hspace*{-1.0cm}
\includegraphics[width=7.8cm]{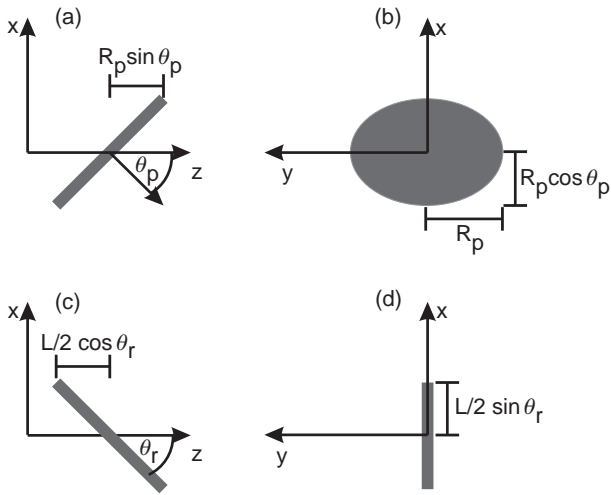}
\end{center}
\caption{Geometries relevant for the determination of the weight functions 
[Eqs.~(\ref{eq14}), (\ref{eq15}), and (\ref{eq18}) - (\ref{eq20})]
for thin circular platelets of radius $R_p$ [(a), (b)] and thin rods of 
length $L$ [(c), (d)]. The angle between the normal of a platelet and the 
$z$-axis is denoted by $\theta_p$ while the angle $\theta_r$ characterizes 
the orientation of a rod with respect to the $z$-axis. Only the projections
of the platelets and rods on the planes of the figures are shown.}
\label{fig1}
\end{figure}

We consider a binary mixture of hard spheres and thin circular platelets of radius 
$R_s$ and $R_p$, respectively. The number density of the centers of mass of 
the platelets at a point ${\bf r}$ with an orientation $\omega_p=(\theta_p,\phi_p)$ 
of the normal of the platelets is denoted by $\rho_p({\bf r},\omega_p)$ while 
$\rho_s({\bf r})$ is the center-of-mass number density of the spheres. The equilibrium 
density profiles of the inhomogeneous mixture under the influence of external 
potentials $V_{ext,s}({\bf r})$ and $V_{ext,p}({\bf r},\omega_p)$ minimize the grand 
potential functional
\begin{eqnarray} \label{eq1}
\lefteqn{\Omega[\rho_s,\rho_p]}\nonumber
\\&=&\int d^3r\,\rho_s({\bf r})
\left[k_BT\left(\ln[\Lambda_s^3\rho_s({\bf r})]-1\right)-
\mu_s+ V_{ext,s}({\bf r})\right]\nonumber
\\&&+\frac{1}{4\pi}\int d^3r\,d\omega_p\,\rho_p({\bf r},\omega_p)
\left[k_BT\left(\ln[\Lambda_p^3\rho_p({\bf r},\omega_p)]-1\right)
\right.\nonumber
\\&&-\left.\mu_p+V_{ext,p}({\bf r},\omega_p)\right]+F_{ex}[\rho_s,\rho_p]\,,
\end{eqnarray}
where $\Lambda_s$, $\Lambda_p$ are the thermal de Broglie wavelengths and 
$\mu_s$, $\mu_p$ are the chemical potentials of the spheres and platelets, 
respectively. The spatial integrals run over the volume V which is accessible 
to the centers of the particles and 
$\int d\omega=\int_0^\pi d\theta\,\int_0^{2\pi} d\phi$. The excess free energy 
functional is obtained by integrating over an excess free energy density,
\begin{equation} \label{eq2}
F_{ex}[\rho_s,\rho_p]=\frac{k_BT}{4\pi}\int d^3r\,d\omega_p\,
\Phi(\{n^{(s)}_\nu,{\bf n}^{(s)}_i,n^{(p)}_\tau,n^{(sp)}\})\,,
\end{equation}
where $\nu=0, 1, 2, 3$, $i=1, 2$, and $\tau=0, 1, 2$. In Eq.~(\ref{eq2}) the 
spatial and angular arguments of the weighted densities $n^{(s)}_\nu$, 
${\bf n}^{(s)}_i$, $n^{(p)}_\tau$, and $n^{(sp)}$ are suppressed in the notation. 
Here we use the following decomposition of the excess free energy density $\Phi$:
\begin{equation} \label{eq3}
\Phi=\Phi_s+\Phi_{sp}\,,
\end{equation}
with \cite{rose:89}
\begin{eqnarray} \label{eq4}
\Phi_s&=&-n_0^{(s)}\ln\left(1-n_3^{(s)}\right)
+\frac{n_1^{(s)}n_2^{(s)}-{\bf n}_{1}^{(s)}\cdot{\bf n}_{2}^{(s)}}{1-n_3^{(s)}}
\nonumber
\\&&+\frac{\left(n_2^{(s)}\right)^3-3n_2^{(s)}\left({\bf n}_{2}^{(s)}\right)^2}
{24\pi\left(1-n_3^{(s)}\right)^2}\,,
\end{eqnarray}
and a new contribution
\begin{eqnarray} \label{eq5}
\Phi_{sp}&=&-n_0^{(p)}\ln\left(1-n_3^{(s)}\right)
+\frac{n_1^{(p)}n^{(sp)}+n_1^{(s)}n_2^{(p)}}{1-n_3^{(s)}}
\nonumber
\\&&+\frac{\pi n_2^{(p)}\left(n_2^{(s)}\right)^2}{64\left(1-n_3^{(s)}\right)^2}\,.
\end{eqnarray}
$\Phi_s$ is the original Rosenfeld excess free energy density \cite{rose:89} for a 
pure hard sphere fluid. (For the subtle issue of the range of validity of the 
Rosenfeld functional at high densities see Refs.~\cite{tara:00,roth:02}.) $\Phi_{sp}$ 
takes sphere-platelet interactions into account up to first order in the number 
density of the platelets (see discussion below). There no is contribution $\Phi_{pp}$
to the excess free energy $\Phi$ in Eq.~(\ref{eq4}) because the platelets are 
considered as non-interacting particles regarding their mutual interactions.
In the present application of density functional theory we concentrate on 
ordering effects induced by a planar hard wall such that the resulting density 
profile of the spheres depends on a single spatial variable $z$ in the direction 
normal to the wall. Hence $\rho_s({\bf r})=\rho_s(z)$ apart from possible surface 
freezing at high densities. Moreover, we assume invariance with respect to 
rotations around the $z$ axis by an angle $\phi_p$, so that 
$\rho_p({\bf r},\omega_p)=\rho_p(z,\theta_p)$, where $\theta_p$ is the angle 
between the normal of a platelet and the $z$-axis (see Fig.~\ref{fig1}). In this 
planar geometry the weighted densities are given 
by 
\begin{eqnarray} \label{eq6}
n^{(s)}_\nu(z)&=&\rho_s(z)\star w^{(s)}_\nu(z)\,,
\\{\bf n}^{(s)}_i(z)&=&\rho_s(z)\star {\bf w}^{(s)}_i(z)\,, \label{eq6a}
\\n^{(p)}_\tau(z,\theta_p)&=&\rho_p(z,\theta_p)\star w^{(p)}_\tau(z,\theta_p)\,,
 \label{eq7}
\\n^{(sp)}(z,\theta_p)&=&\rho_s(z)\star w^{(sp)}(z,\theta_p)\,,
 \label{eq8}
\end{eqnarray}
where the asterics $\star$ denotes the 
spatial convolution: $g(z)\star h(z)=\int dz_1\, g(z_1) h(z-z_1)\equiv g\star h$. 
Note that $n^{(s)}_\nu$, ${\bf n}^{(s)}_i$, and  $n^{(p)}_\tau$ are weighted 
densities which involve only variables of either species, while $n^{(sp)}$ is 
a convolution of the sphere density with an  orientation-dependent weight function,
combining characteristics of both species. The weight functions of the Rosenfeld 
excess free energy density read
\begin{eqnarray} \label{eq9}
w^{(s)}_0(z)&=&\frac{\Theta(R_s-|z|)}{2R_s}\,,
\\w^{(s)}_1(z)&=&\frac{\Theta(R_s-|z|)}{2}=\frac{w^{(s)}_2(z)}{4\pi R_s}\,,
 \label{eq10}
\\w^{(s)}_3(z)&=&\pi(R_s^2-z^2)\Theta(R_s-|z|)\,,
 \label{eq11}
\\{\bf w}^{(s)}_1(z)&=&\frac{z\Theta(R_s-|z|){\bf e}_z}{2R_s}
=\frac{{\bf w}^{(s)}_2(z)}{4\pi R_s}\,,
 \label{eq12}
\end{eqnarray}
where ${\bf e}_z$ is the unit vector pointing along the z axis and $\Theta(z)$
is the Heaviside step function. The Mayer function $f_{ss}$(z) of the interaction 
potential between two hard spheres is obtained through
\begin{equation} \label{eq13}
-\frac{f_{ss}(z)}{2}=w^{(s)}_0\star w^{(s)}_3+
w^{(s)}_1\star w^{(s)}_2-{\bf w}^{(s)}_{1}\star {\bf w}^{(s)}_{2}\,.
\end{equation}
The Mayer function equals $-1$ if the spheres intersect or touch each other and 
is zero otherwise. The remaining weight functions can be expressed as
(see discussion below)
\begin{eqnarray} \label{eq14}
w^{(p)}_0(z,\theta_p)&=&\frac{\delta(R_p\sin\theta_p-|z|)}{2}\,,
\\w^{(p)}_1(z,\theta_p)&=&\frac{\pi\Theta(R_p\sin\theta_p-|z|)}{8\sin\theta_p}
=\frac{w^{(p)}_2(z,\theta_p)}{8R_p}\!,
 \label{eq15}
\end{eqnarray}
and
\begin{equation} \label{eq16}
w^{(sp)}(z,\theta_p)=\left\{
\begin{array}{ll}
8\sqrt{R_s^2\cos^2\theta_p-z^2}
\\+8z\left[\arcsin\left(\frac{z\tan\theta_p}{\sqrt{R_s^2-z^2}}\right)\right]
\sin\theta_p,
\\&\hspace*{-2.8cm}|z|<R_s\cos\theta_p\\
4\pi |z|\sin\theta_p, &\hspace*{-2.8cm}R_s\cos\theta_p \le |z| \le R_s\,.
\end{array}
\right.
\end{equation}
These weight functions allow one to generate the Mayer function $f_{sp}(z)$ of 
the interaction potential between a hard sphere and a thin platelet through 
\begin{equation} \label{eq17}
-f_{sp}(z,\theta_p)=w^{(p)}_0\star w^{(s)}_3+
w^{(p)}_1\star w^{(sp)}_2+w^{(s)}_1\star w^{(p)}_2\,.
\end{equation}
Equations (\ref{eq1}) - (\ref{eq17}) completely specify the density functional 
theory for the system under consideration.

Before studying the binary mixture of spheres and thin platelets in the bulk and near 
a hard wall it is instructive to compare the fundamental measure theory with the 
one which has been developed recently for a mixture of hard spheres and thin rods
\cite{schm:01a,schm:01b,schm:02,brad:02a}. For thin rods of length $L$ the weight 
functions corresponding to Eqs.~(\ref{eq14}) - (\ref{eq16}) are given by 
\begin{eqnarray} \label{eq18}
w^{(r)}_0(z,\theta_r)&=&\frac{\delta\left(\frac{L}{2}\cos\theta_r-|z|\right)}{2}\,,
\\w^{(r)}_1(z,\theta_r)&=&\frac{\Theta\left(\frac{L}{2}\cos\theta_r-|z|\right)}
{4\cos\theta_r}\,,
 \label{eq19}
\\w^{(r)}_2(z,\theta_r)&=&0\,,
 \label{eq20}
 \end{eqnarray}
and
\begin{equation} \label{eq21}
w^{(sr)}(z,\theta_r)=\left\{
\begin{array}{ll}
8\sqrt{R_s^2\sin^2\theta_r-z^2}
\\+8z\left[\arcsin\left(\frac{z\cot\theta_r}{\sqrt{R_s^2-z^2}}\right)\right]
\cos\theta_r,
\\&\hspace*{-2.8cm}|z|<R_s\sin\theta_r
\\4\pi |z|\cos\theta_r, &\hspace*{-2.8cm}R_s\sin\theta_r \le |z| \le R_s\,.
\end{array}
\right.
\end{equation}
Here $\theta_r$ is the angle between the rod and the $z$ axis (see 
Fig.~\ref{fig1}). The Mayer function $f_{sr}(z)$ of the interaction potential 
between a hard sphere and a thin rod is generated through
\begin{equation} \label{eq22}
-f_{sr}(z,\theta_r)=w^{(r)}_0\star w^{(s)}_3+w^{(r)}_1\star w^{(sr)}_2\,.
\end{equation}
The weight functions are linked with a geometrical representation of the 
particles which is given in terms of fundamental measures defined as 
$\zeta^{(j)}_\lambda=\int dz\,w^{(j)}_\lambda$, where $j=s, p, r$ labels 
the species, and $\lambda=0,1,2,3$ corresponds to the Euler characteristic,
integral mean curvature, surface, and volume \cite{rose:94} of the particles. 
For spheres 
$\zeta^{(s)}_0=1$, $\zeta^{(s)}_1=R_s$, $\zeta^{(s)}_2=4\pi R_s^2$, and
$\zeta^{(s)}_3=4\pi R_s^3/3$, whereas for thin platelets the volume
is very small and $\zeta^{(p)}_0=1$, $\zeta^{(p)}_1=\pi R_p/4$, and 
$\zeta^{(p)}_2=2\pi R_p^2$. In the case of thin rods both the volume 
and the surface area are very small and $\zeta^{(r)}_0=1$, $\zeta^{(r)}_1=L/4$.
For comparison we note that the integral mean curvature of a particle
can be obtained from the general relation 
$\zeta_1=\int d\sigma\,(1/R_1+1/R_2)/(8\pi)$, where $R_1$ and $R_2$ 
are the principle radii of curvature at the point $\sigma$ on the 
surface and $d\sigma$ is a surface element. The evaluation of this 
integral is trivial for thin rods and is documented for thin platelets 
in the appendix of Ref.~\cite{bate:99}.

From their side view, platelets may be regarded as rods 
(see Figs.~\ref{fig1} (a) and (c)) and from their top view (in the direction 
of the normal to face) as two-dimensional spheres. Therefore the functional 
forms of the weight functions $w^{(p)}_0(z)$, $w^{(r)}_0(z)$ 
[Eqs.~(\ref{eq14}) and (\ref{eq18})] and $w^{(p)}_1(z)$, $w^{(r)}_1(z)$
[Eqs.~(\ref{eq15}) and (\ref{eq19})] are similar, while the weight function 
$w^{(p)}_2(z)$ [Eq.~(\ref{eq15})] takes into account the surface of the 
platelets (see Fig.~\ref{fig1} (b)). 

\begin{figure}[t]
\begin{center}
\hspace*{-0.5cm}
\includegraphics[width=7.8cm]{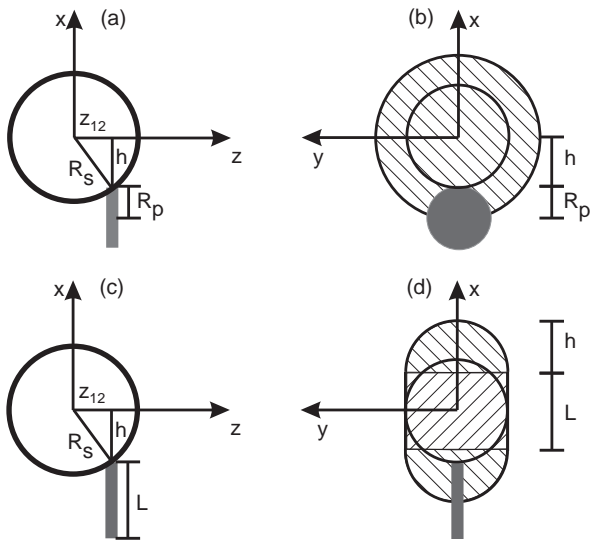}
\end{center}
\caption{Illustrations of the steric interactions of hard spheres with 
thin hard platelets and rods. (a) [(c)] Schematic side view 
of a sphere of radius $R_s$ and a platelet of radius 
$R_p$ [rod of length $L$]. The normal of the platelet is parallel to the 
$z$-axis while the rod is oriented parallel to the $x$-axis. Only the projections 
of the particles on the planes of the figures are shown.
(b) [(d)] Due to the steric interaction, the center of mass of the platelet [rod] 
is excluded from the hatched region surrounding the sphere in the 
$x-y$ plane. Here the planes of the figures are located at $z=z_{12}$. 
From these figures the area $A_{sp}(z_{12})=\pi (h+R_p)^2$ 
[$A_{sr}(z_{12})=\pi h^2+2Lh$] with $h=\sqrt{R_s^2-z_{12}^2}$ can be inferred
(see Eq.~(\ref{eq24}) [(\ref{eq26})]).}
\label{fig2}
\end{figure}

In order to express the sphere-platelet and sphere-rod Mayer functions 
[Eqs.~(\ref{eq17}) and  (\ref{eq22})] in terms of spatial convolution decompositions,
additional weight functions $w^{(sp)}(z,\theta_p)$ and $w^{(sr)}(z,\theta_r)$ 
have to be introduced. These weight functions contain informations about both 
species of the binary mixtures. Since the explicit expression for the additional 
weight function [Eq.~(\ref{eq21})] for the sphere-rod mixture is amply discussed 
in Ref.~\cite{brad:02a} we do not document the derivation of the expression for 
the corresponding weight function $w^{(sp)}(z,\theta_p)$ [Eq.~(\ref{eq16})] 
for the sphere-platelet mixture but analyze the cases $\theta_p=0$ and 
$\theta_r=\pi/2$ in more detail. In these limits the weight functions reduce to 
$w^{(sp)}(z,\theta_p=0)=w^{(sr)}(z,\theta_r=\pi/2)=8\sqrt{R_s^2-z^2}\,\Theta(R_s-|z|)$.
Figure \ref{fig2} displays schematic illustrations of the support of the 
sphere-platelet and sphere-rod Mayer functions which can be calculated analytically 
from Eqs.~(\ref{eq17}) and (\ref{eq22}):
\begin{eqnarray} \label{eq23}
-f_{sp}(z,\theta_p=0)&=&A_{sp}(z)\Theta(R_s-|z|)\,,
\\A_{sp}(z)&=&\pi\left(\sqrt{R_s^2-z^2}+R_p\right)^2\,,
 \label{eq24}
\end{eqnarray}
and
\begin{eqnarray} \label{eq25}
-f_{sr}(z,\theta_r=\frac{\pi}{2})&=&A_{sr}(z)\Theta(R_s-|z|)\,,
\\A_{sr}(z)&=&\pi(R_s^2-z^2)+2L\sqrt{R_s^2-z^2}\,.
 \label{eq26}
\end{eqnarray}
Due to the steric interaction, the center of mass of a platelet [rod] is excluded 
from an area in the $x-y$ plane of size $A_{sp}(z_{12})$ [$A_{sr}(z_{12})$] 
surrounding a sphere, where $z_{12}$ is the distance 
along the $z$ axis between the center of mass of the platelet [rod] and the 
sphere. Moreover it is apparent from the figure that the radius 
$\sqrt{R_s^2-z^2}$ and hence the weight function $w^{(sp)}(z,0)$
[$w^{(sr)}(z,\pi/2)$] are characteristic of how a sphere looks from the 
viewpoint of a platelet [rod].

\section{Bulk phase diagram} 
Based on the density functional given by Eqs.~(\ref{eq1}) - (\ref{eq17}) and as 
a prerequisite for our wall-liquid interface study we first study 
the homogeneous bulk fluid with $V_{ext,s}({\bf r})=0$ and 
$V_{ext,p}({\bf r},\omega_p)=0$ in a macroscopic volume $V$. 
In this case the equilibrium density profiles are constant
[$\rho_s({\bf r})=\rho_s$ and $\rho_p({\bf r},\omega_p)=\rho_p$] and the 
Euler-Lagrange equations resulting from the stationary conditions are given 
by $\partial \Omega[\rho_s,\rho_p]/\partial \rho_s=0$ and 
$\partial \Omega[\rho_s,\rho_p]/\partial \rho_p=0$. In the present study we 
restrict our attention to platelet number densities $\rho_p R_p^3 \le 0.2$
for which the pure platelet fluid and also the mixture are in the isotropic 
phase and hence $\rho_s$ is independent of the orientation of the platelets. 
For comparison, the isotropic-nematic phase transition of the pure platelet fluid 
is first order with coexistence densities $\rho_{pI}R_p^3=0.46$ and 
$\rho_{pN}R_p^3=0.5$ according to a computer simulation \cite{bate:99}. The excess 
free energy density per volume can be expressed as
\begin{equation} \label{eq27}
\frac{F_{ex}}{k_BTV}=\Phi_{s,b}-\rho_p\,\ln\alpha\,,
\end{equation}
with
\begin{eqnarray} \label{eq28}
\lefteqn{\frac{\alpha}{1-\eta_s}}\nonumber
\\&=&\exp\left(-\frac{\left(\pi^2 R_s^2 R_p+2\pi R_p^2 R_s\right)\rho_s}{1-\eta_s}
-\frac{\pi^4 R_s^4 R_p^2\rho_s^2}{2(1-\eta_s)^2}\right),\nonumber
\\
\end{eqnarray}
where $\eta_s=4\pi R_s^3 \rho_s/3$ is the packing fraction of the spheres and 
$\Phi_{s,b}$ is the excess free energy density of a pure hard-sphere fluid. 
$\alpha$ can be considered as the free-volume fraction, i.e., the relative amount 
of the volume $V$ that is accessible to the platelets. Besides providing access 
to inhomogeneous density distributions the above geometry-based density 
functional theory for mixtures of hard spheres and platelets 
offers in addition a systematic approach to calculate the work $W=-k_BT\ln \alpha$ 
required to insert a platelet into a solution of spheres. The expression 
for the free-volume fraction $\alpha$ in Eq.~(\ref{eq28}) is equivalent to the result 
from a recent scaled-particle approach \cite{over:04}. For comparison we note that 
the corresponding free-volume fraction for thin rods is given by
\begin{equation} \label{eq29}
\frac{\alpha}{1-\eta_s}=\exp\left(-\frac{\pi L R_s^2\rho_s}{1-\eta_s}\right)\,.
\end{equation}
From the bulk grand potential function all thermodynamic quantities can be calculated. 
Equating the pressure and the chemical potentials of both species in both phases 
yields the coexisting densities. 

Figure \ref{fig3} (a) displays the calculated phase diagram for a binary mixture 
of spheres and thin platelets for size ratio $R_s/R_p=8/3$ as a function of the 
chemical potential $\mu_p$ of the platelets and the number density $\rho_s$
of the spheres. The tie-lines are horizontal because of the equality of $\mu_p$ of the 
coexisting phases. The binodal for coexisting states is shown, where a sphere-rich 
and a platelet-poor liquid phase coexists with a sphere-poor and a platelet-rich liquid 
phase. The coexistence region is bounded by a lower critical point below which only a 
single stable phase is found. For convenience we have introduced the dimensionless 
variable $\mu_p^\star\equiv \mu_p-k_BT\ln(\Lambda_p^3/R_p^3)$, and dropped the star in 
order to avoid a clumsy notation. Figure \ref{fig3} (b) displays an alternative 
representation of the phase diagram in terms of the number densities of both species. 
The figure illustrates the fractionation of both spheres and platelets due to the 
phase transition. Upon increasing the size ratio the critical point shifts to larger 
densities of the spheres. 

The dotted lines in Fig.~\ref{fig3} represent the binodal as calculated without the 
term proportional to $\rho_s^2$ in parenthesis on the r.h.s. of Eq.~(\ref{eq28}) which 
is equivalent to considering a binary mixture of spheres and thin rods of length
$L=\pi R_p+2 R_p^2/R_s=1.46\, R_s$ as is apparent from a comparison of Eqs.~(\ref{eq28}) 
and (\ref{eq29}). The phase boundaries and the lower critical point are shifted to  
smaller values of $\rho_s$. In other words, the number density of the spheres in the 
sphere-poor phase is increased by taking into account the last term in parenthesis 
in Eq.~(\ref{eq28}).

\section{Binary sphere-platelet mixture near a planar hard wall} 
The density profiles of both components of the binary mixture of hard spheres and thin 
platelets close to a planar hard wall are obtained by a numerical minimization of 
the grand potential functional (\ref{eq1}) with the excess free energy functional 
given by Eq.~(\ref{eq2}). 
\begin{figure}[t]
\begin{center}
\hspace*{-0.5cm}
\includegraphics[width=8cm]{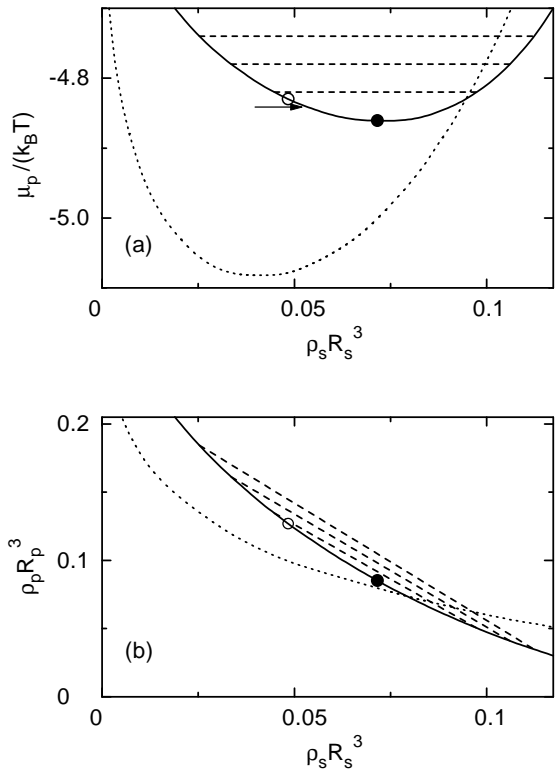}
\end{center}
\vspace*{-1.0cm}
\caption{(a) Bulk and surface phase diagrams of binary mixtures of hard spheres of radius 
$R_s$ and thin hard platelets of radius $R_p=3R_s/8$ as a function of the chemical 
potential of the platelets $\mu_p$ and the number density  of the spheres $\rho_s$. 
The straight dashed lines are tie-lines illustrating liquid-liquid phase 
coexistence. (b) Phase diagram of the same fluid in the density-density plane, where 
$\rho_p$ is the number density of the platelets. In (a) and (b) the solid and open 
circles denote the bulk critical point and the wetting transition point, respectively. 
Between the wetting transition point and the critical point the sphere-rich liquid 
phase completely wets the interface between the hard wall and the sphere-poor liquid phase.
The dotted curve is the binodal as calculated without the last term in parenthesis 
on the r.h.s. of Eq.~(\ref{eq28}) which is equivalent to the bulk phase diagram of a 
binary mixture of spheres of radius $R_s$ and thin rods of length $L=1.46\, R_s$ 
[see the main text]. In Fig.~\ref{fig4} density profiles near a hard wall are shown 
along the thermodynamic path indicated by the arrow at $\mu_p/(k_BT)=-4.85$ in (a);
in (b) this path would run parallel to the dashed tie-lines.}
\label{fig3}
\end{figure}
We fix the chemical potential $\mu_p$ of the platelets and 
approach the bulk phase boundary from the sphere-poor side. Upon decreasing $\mu_p$ 
the adsorption behavior changes qualitatively, and it is worthwhile to distinguish 
the following two cases. For $\mu_p/(k_BT)>-4.84$ we find that the wall is only 
partially wet by the spheres. The layer thickness of the sphere-rich phase forming
close to the wall increases continuously, but remains finite at coexistence. 
For $\mu_p/(k_BT)<-4.84$ we observe complete wetting. The transition to complete 
wetting appears to be first order because the excess adsorptions
$\Gamma_l=R_l^2\int_0^\infty d z\,(\rho_l(z)-\rho_l(\infty))$ with $l=p, r$, 
calculated along the coexistence curve jump to a macroscopic value upon approaching 
the wetting transition point. 
\begin{figure}[t]
\begin{center}
\hspace*{-0.5cm}
\includegraphics[width=8cm]{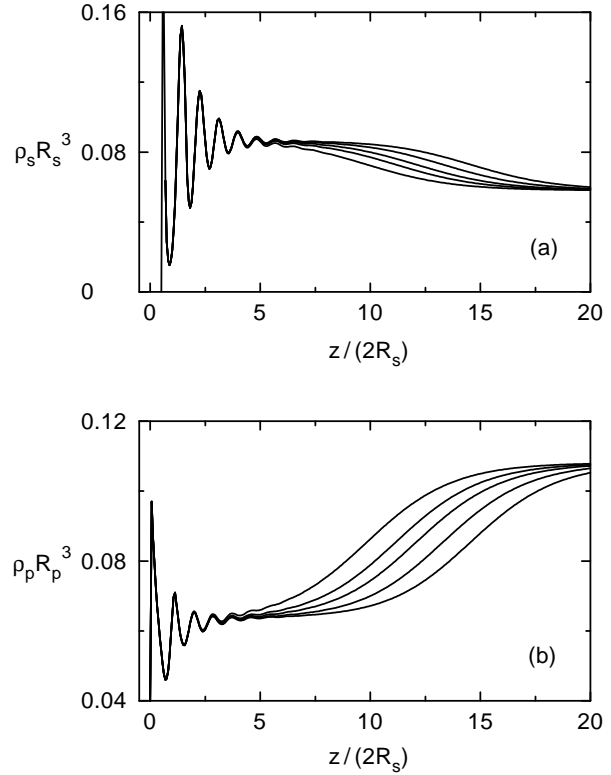}
\end{center}
\vspace*{-1.0cm}
\caption{Equilibrium density profiles of hard spheres of radius $R_s$ (a)
and thin hard platelets of radius $R_p=3 R_s/8$ (b) in contact with a planar 
hard wall at $z=0$ as the bulk phase boundary is approached along the path 
indicated by the arrow in Fig.~\ref{fig3} (a). The chemical potentials of the spheres 
are $\mu_s/(k_BT)=25.82862, 25.82866, 25.8287, 25.82874, 25.82877$ corresponding to 
$\rho_s R_s^3=0.058057, 0.058069, 0.058082, 0.058094, 0.058104$
(from left to right) 
where the chemical potential at bulk coexistence is $\mu_s/(k_BT)=25.8287774$ so that 
$\rho_s R_s^3=0.058107$.
In (a) $\rho_s(z<R_s)=0$ and in (b) $\rho_p(z<0)=0$; the contact values at $z=R_s$ 
and $z=0$, respectively, are not shown on the present scales.}
\label{fig4}
\end{figure}
Figure \ref{fig4} (a) displays the sphere 
density profiles at $\mu_p/(k_BT)=-4.85$ signalling the growth of a thick layer of 
sphere liquid at the wall. The corresponding platelet profiles are shown in 
Fig.~\ref{fig4} (b) and indicate how the platelets become more depleted as the 
sphere-rich layer grows. Upon approaching the chemical potential of the spheres 
at bulk coexistence $\mu_s/(k_BT)=25.8287774$, the calculated density profiles at 
the liquid-liquid interface become virtually indistinguishable from the ones of 
the free liquid-liquid interface between coexisting bulk phases, and the layer 
thickness diverges logarithmically, as expected for the case of complete wetting
in systems governed by short-ranged forces. 
With increasing chemical potential of the platelets, and hence increasing distance 
to the critical point [see Fig.~\ref{fig3} (a)], the interface becomes sharper, i.e., 
it crosses over from one to the other limiting bulk value over a shorter distance.
The wavelength $\lambda=1.72\,\,R_s$ of the oscillations of the density profiles 
close to the wall reflects the size of the spheres.

Similar to recent studies of wetting in sphere-polymer 
\cite{brad:02b,dijk:02} and sphere-rod \cite{roth:03} mixtures, we have not been able 
to numerically resolve the prewetting line which should emerge tangentially from the 
coexistence curve at the wetting transition. In contrast to those studies there are 
no layering transitions in the partial wetting regime for the sphere-platelet mixture. 
This holds also for the aforementioned toy model without the last term in Eq.~(\ref{eq4}) 
which exhibits the same bulk phase diagram as the corresponding sphere-rod mixture 
(see the dotted curves in Fig.~\ref{fig3}). Taking into account
the weight function $w_2^{(p)}(z,\theta_p)$ [Eq.~(\ref{eq15})] leads to wetting 
phenomena of sphere-platelet mixtures which are different from those of sphere-rod 
mixtures even if the bulk phase diagrams of both systems are identical. 
Finally, we note that the wetting behavior discussed above remains unchanged upon 
increasing the size ratio of the platelets and the spheres (e.g., $R_p=R_s/2$).

\section{Summary} 
We have developed a geometry-based density functional theory for fluids 
consisting of hard spheres and thin platelets. The bulk and surface phase 
diagram and the density profiles near a planar hard wall are determined 
numerically with the following main results.

(1) Figure \ref{fig1} illustrates that from their side view thin platelets may 
be regarded as thin rods and from their top view as two-dimensional spheres. 
On the basis of this consideration we have shown that the geometry-based density 
functional theory developed
 for binary mixture of hard spheres and thin rods
\cite{schm:01a,schm:01b,schm:02,brad:02a} 
can be consistently extended to the more challenging problem of hard spheres mixed 
with thin hard platelets by introducing an additional weight function which 
characterizes the surface of a platelet. The volume accessible to a thin platelet 
of radius $R_p$ in the presence of a sphere is smaller than the corresponding 
one of a thin rod of length $L=2R_p$ because of the extended surface of the platelet 
(Fig.~\ref{fig2}).

(2) The bulk phase diagram exhibits two-phase coexistence between sphere-rich 
and sphere-poor phases which is bounded by a lower critical point below which 
a single stable phase is found (Fig.~\ref{fig3}). The phase boundaries and the 
critical point of the corresponding phase diagram for a binary mixture of hard 
spheres and thin rods are shifted to  smaller values of the density of the 
spheres due to smaller intermolecular interactions between thin rods and spheres 
as compared with those between thin platelets and spheres.

(3) For the mixture near a planar hard wall, a first-order wetting transition 
by the sphere-rich phase occurs. In the partial wetting regime no 
layering transitions are found in contrast to recent studies on binary 
mixtures of spheres and non-adsorbing polymers or thin hard rods.

We have focused on the case of non-interacting platelets as regards their mutual 
interactions, which constitutes a minimal model for non-spherical particles
with non-vanishing surface area. With increasing density of the platelets, 
interactions between platelets must be included \cite{harn:01,harn:02}. The consistent 
treatment of these nontrivial platelet-platelet interactions within a geometry-based 
density functional theory remains as a challenge.

\acknowledgments
The authors thank R. Roth for useful discussions.

\end{document}